\documentclass[sigplan,screen]{acmart}
\AtBeginDocument{%
  }

\setcopyright{acmlicensed}
\copyrightyear{2025} 
\acmYear{2025} 
\setcopyright{acmlicensed}\acmConference[ICMR '25]{Proceedings of the 2025
 International Conference on Multimedia Retrieval}{June 30-July 3,
 2025}{Chicago, IL, USA}
 \acmBooktitle{Proceedings of the 2025 International Conference on Multimedia
 Retrieval (ICMR '25), June 30-July 3, 2025, Chicago, IL, USA}
 \acmDOI{10.1145/3731715.3733366}
 \acmISBN{979-8-4007-1877-9/2025/06}
\usepackage{multirow}
\usepackage{times}  
\usepackage{helvet}  
\usepackage{courier}  

\usepackage{graphicx} 
\usepackage{newfloat}
\usepackage{listings}
\usepackage{amsmath}   
\usepackage{bm}     
\usepackage{mathtools}
\DeclarePairedDelimiterX{\infdivx}[2]{(}{)}{%
  #1\;\delimsize\|\;#2%
}
\newcommand{\infdiv}{D_\textrm{KL}\infdivx}

\begin{document}

\title{Inter - Diffusion Generation Model of Speakers and Listeners for Effective Communication}

\author{Jinhe Huang}
\authornote{equal contribution}
\email{19217124@njau.edu.cn}
\affiliation{%
  \institution{Nanjing Agriculture University}
  \country{China}
}

\author{Yongkang Cheng}
\authornotemark[1]
\authornote{project leader}
\email{19217105@njau.edu.cn}
\affiliation{%
  \institution{Nanjing Agriculture University}
  \country{China}}

\author{Minghang Yu*}
\email{1431605505@qq.com}
\affiliation{%
  \institution{Northwest A\&F University}
  \country{China}}

\author{Gaoge Han}
\email{hangaoge@nwafu.edu.cn}
\affiliation{%
  \institution{City University of Hong Kong}
  \country{China}}

\author{Jinwei Li}
\email{lijinwei231@mails.ucas.ac.cn}
\affiliation{%
  \institution{University of Chinese Academy of Sciences}
  \country{China}}

\author{Jing Zhang}
\email{m17863974365@163.com}
\affiliation{%
  \institution{University of Chinese Academy of Sciences}
  \country{China}}

\author{Shilei Wang}
\email{shileiwang@mail.nwpu.edu.cn}
\affiliation{%
  \institution{Northwestern Polytechnical University}
  \country{China}}

\author{Xingjian Gu }
\authornote{corresponding author}
\email{guxingjian@njau.edu.cn}
\affiliation{%
  \institution{Nanjing Agriculture University}
  \country{China}}
  






\renewcommand{\shortauthors}{Huang, et al.}

\begin{abstract}
  Full-body gestures play a pivotal role in natural interactions and are crucial for achieving effective communication. Nevertheless, most existing studies primarily focus on the gesture generation of speakers, overlooking the vital role of listeners in the interaction process and failing to fully explore the dynamic interaction between them. This paper innovatively proposes an Inter-Diffusion Generation Model of Speakers and Listeners for Effective Communication.
For the first time, we integrate the full-body gestures of listeners into the generation framework. By devising a novel inter-diffusion mechanism, this model can accurately capture the complex interaction patterns between speakers and listeners during communication. In the model construction process, based on the advanced diffusion model architecture, we innovatively introduce interaction conditions and the GAN model to increase the denoising step size. As a result, when generating gesture sequences, the model can not only dynamically generate based on the speaker's speech information but also respond in realtime to the listener's feedback, enabling synergistic interaction between the two. Abundant experimental results demonstrate that compared with the current state-of-the-art gesture generation methods, the model we proposed has achieved remarkable improvements in the naturalness, coherence, and speech-gesture synchronization of the generated gestures. In the subjective evaluation experiments, users highly praised the generated interaction scenarios, believing that they are closer to real life human communication situations. Objective index evaluations also show that our model outperforms the baseline methods in multiple key indicators, providing more powerful support for effective communication.
\end{abstract}





\maketitle


\section{Introduction}
In the vast and evolving landscape of human communication research, the generation of natural looking body gestures has been a long standing and challenging pursuit, particularly in the burgeoning fields of virtual reality, gaming, and filmmaking. These non-verbal cues are far more than just supplementary elements; they are the lifeblood of immersive and realistic interactions. By seamlessly integrating with spoken language, body gestures can convey a wealth of information, including emotions, attitudes, and intentions, thereby enriching the overall communication experience.

The majority of existing research efforts have been predominantly fixated on generating gestures for speakers, overlooking the equally crucial role of listeners in the communication ecosystem. Listeners are not passive bystanders in the communication process. Their full body gestures actively participate in the communication feedback loop. For instance, a listener's nodding can signal understanding and encourage the speaker to continue, while a furrowed brow or a puzzled expression might prompt the speaker to clarify their statement. In real world interactions, these non-verbal cues from listeners can have a profound impact on the speaker's subsequent actions, speech patterns, and even the direction of the conversation. The lack of consideration for the listener's global gestures in current research has thus created a disconnect between the generated virtual interactions and the rich, dynamic nature of real life communication. Virtual communication scenarios often feel stilted and unrealistic because they fail to capture the fluid back and forth between speakers and listeners.

Another significant hurdle in this research area is the scarcity of data related to listeners. Most existing datasets are primarily centered around speaker gestures, leaving a dearth of information on how listeners express themselves non-verbally during communication. This lack of listener-focused data severely restricts the development of models that can accurately represent the complete communication spectrum. Fortunately, we discovered that the TWH dataset contains a substantial amount of listener related data. Complemented by the zeroegg dataset, we were able to pair them up to create a comprehensive dataset for studying the global interactive gestures in communication. These two datasets, when combined, provide a rich source of information on the back and forth non-verbal communication between speakers and listeners. We utilized this assembled dataset to train our model, enabling it to learn the complex patterns of interaction between the two parties.

This glaring deficiency in existing research forms the core motivation for our work. We aim to revolutionize the field of gesture generation by addressing these long - standing issues. In this paper, we present a groundbreaking contribution: for the first time, we introduce the generation of the listener's global gestures into the framework. Our Inter-Diffusion Generation Model of Speakers and Listeners for Effective Communication is meticulously designed to bridge this substantial gap. Our model is built upon a novel Inter-diffusion mechanism. This mechanism serves as the cornerstone of our approach, enabling the model to effectively capture the intricate and dynamic interaction patterns between speakers and listeners. By leveraging an advanced diffusion model architecture, we introduce innovative interaction conditions. These conditions endow the model with the remarkable ability to generate gestures for speakers based on their speech content. Simultaneously, it can generate context appropriate global gestures for listeners in response, creating a seamless two-way interaction. This two-way interaction framework is a departure from traditional models, as it mirrors the natural ebb and flow of real world communication, resulting in a more natural and interactive communication environment.

When we incorporate listeners into the generation model, distinct and remarkable visual effects emerge. In virtual communication scenes, listeners' body postures now exhibit natural inclinations towards the speakers, as if they are fully engaged in the conversation. Their head movements, such as gentle nods or slight tilts, synchronize with the rhythm of the speaker's speech, giving the impression of active listening. These visual cues, combined with the generated speaker gestures, create a more immersive and life - like communication environment. The once - static virtual listeners now seem to be living, breathing participants in the conversation, making the entire virtual communication experience far more engaging and realistic.

To validate the effectiveness of our model, we conducted a series of comprehensive experiments. The results unequivocally demonstrate the superiority of our model over existing methods. It generates more authentic and context-aware gesture sequences for both speakers and listeners. In virtual communication scenarios, the gestures generated by our model are more in tune with the ongoing conversation, enhancing the overall quality of these interactions. This research not only enriches the body of knowledge in gesture generation but also holds great promise for a wide range of applications. From improving the user experience in virtual reality and gaming to enhancing the authenticity of characters in film and television, our model can be a game changer in any application that requires realistic and interactive communication simulations.

\section{Related Work}
\noindent\textbf{Speaker Generation Model} 
Co-speech gesture generation is a complex task that demands the integration of speech melody, semantics, and gesture motion. In the early days, data - driven methods, as seen in ~\cite{liu2022audio,habibie2021learning, cheng2024expgest, yang2024freetalker}, learned directly from human demonstrations. These models were limited in motion diversity, often producing repetitive gestures as they were overly reliant on the training data. They mainly focused on generating gestures for speakers, overlooking the dynamic nature of real - world communication.
Subsequent research, such as ~\cite{habibie2021learning,yi2023generating,xie2022vector}, strived to create more diverse and expressive gestures. These studies explored innovative ways to represent the latent space of gestures and speech, yet they still remained within the framework of only considering speaker - side generation. Some efforts ~\cite{yang2023diffusestylegesture,yang2023unifiedgesture,ahuja2020style,ao2023gesturediffuclip} centered on training unified models for multiple speakers. They either embedded speaker styles in the latent space or utilized style-transfer techniques. Meanwhile, ~\cite{zhou2022gesturemaster,habibie2022motion} employed motion matching, but this approach was burdened with complex matching rules. Notably, most existing audio - driven gesture generation models only account for speakers. They lack the ability to capture the feedback and non - verbal cues from listeners, which are vital in natural communication. Listeners' gestures and body language can influence the speaker's subsequent speech and actions, creating a dynamic interaction loop. Our work is groundbreaking as we introduce the generation of listener's global gestures for the first time. By incorporating listeners into the model, we can create a more realistic and interactive communication environment, filling a significant gap in the current research. Despite challenges, audio-driven animation, including co-speech gesture generation, remains a popular research area due to its wide applications in VR, gaming, and film.

\noindent\textbf{Diffusion Generative Models} 
Diffusion Generative Models~\cite{zhao2023modiff, wang2022diffusion} have made significant progress in human motion generation. Motion Diffuse first applied them to text-conditioned motion~\cite{zhang2022motiondiffuse, han2025reindiffuse}, enabling fine-grained body part control. MDM~\cite{tevet2022human} was a milestone, manipulating motion based on text inputs. GestureClip~\cite{ao2023gesturediffuclip} used an attention mechanism for speech matched results. However, these methods have limitations. They operate within a single speaker paradigm, relying on independent diffusion frameworks without interactive structures. This means they ignore the role of listeners, which is crucial in real world communication.

In contrast, our work innovatively proposes an interactive diffusion model. This model takes into account both speakers and listeners, filling the gap in the existing single speaker-centered frameworks. By enabling dynamic interaction between the two, it can generate more realistic and context-aware motion sequences, bringing the motion generation closer to real world communication scenarios.

\section{Method}

\begin{figure}
  \begin{center}
    \includegraphics[width=1\linewidth]{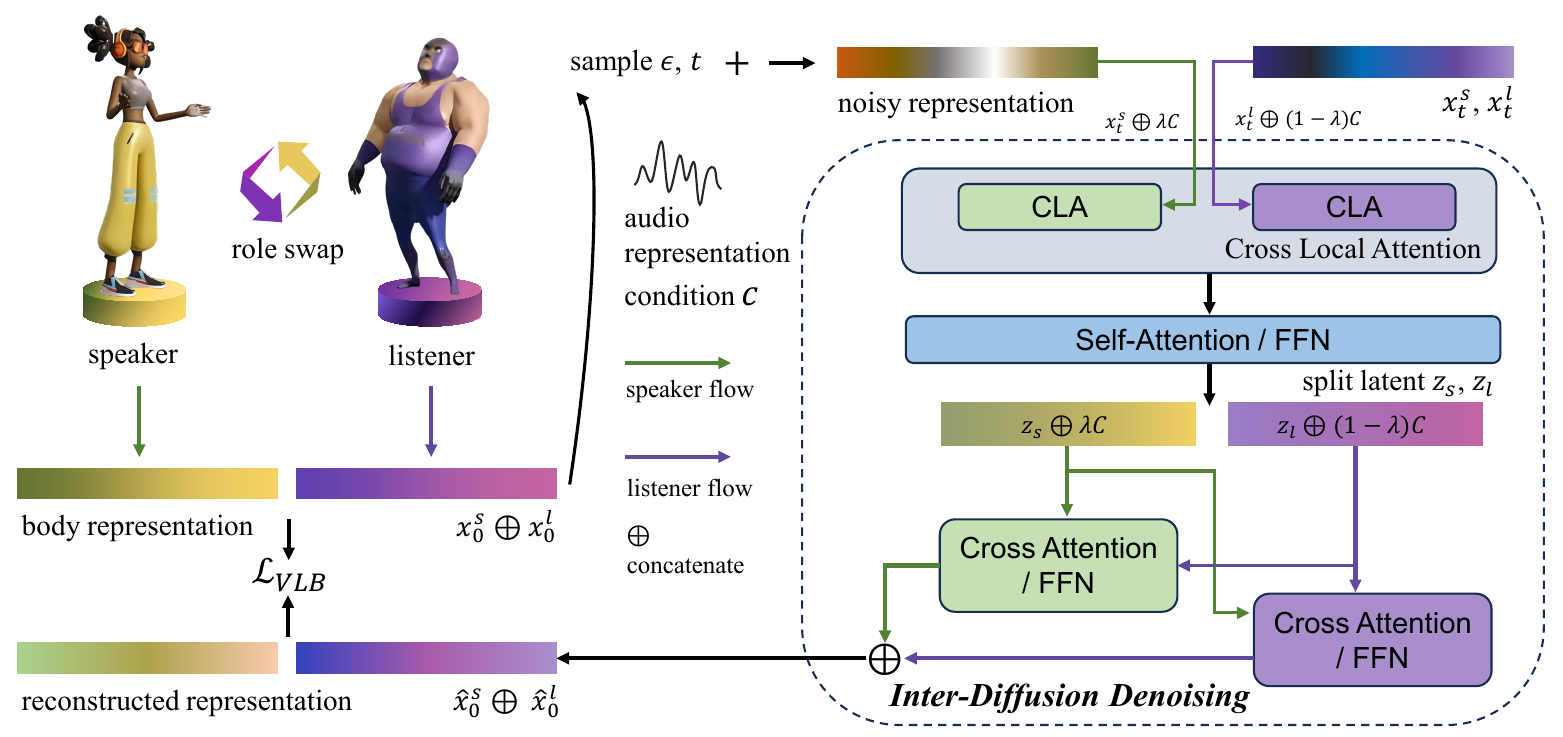}
  \end{center}
  \vspace{-10pt}
  \caption{Overview training process of our Inter-Diffusion.}
  \Description[<short description>]{}
  \vspace{-10pt}
  \label{fig:pipe}
\end{figure}

\subsection{System Overview}

Our framework aims to address the lack of conversational priors in previous methods, which only modeled isolated speech of the speaker, by introducing vocal interactions between the speaker and listener during the dialogue. To maintain consistency and ensure a fair comparison with prior approaches, we continue to employ the MDM framework for training our model. MDM is a variance-preserving diffusion model whose training objective is equivalent to optimizing the variational lower bound loss in DDPM\cite{ho2020denoising}. In simpler terms, when no conditions are introduced, for a sample \( x_t \) drawn from the noise distribution \( q(x_t | x_0) \), the denoiser directly predicts the original sample \( \hat{x}_0 \) at that timestep. The \( L2 \)-loss between \( \hat{x}_0 \) and \( x_0 \sim p_{data}(x) \) sampled from the data distribution is then used to estimate the log-gradient (score)\cite{song2020score} under this noise distribution\cite{ho2022classifier}.

When a conditional diffusion model is required, the condition \( C \) is incorporated as an additional vector into the intermediate variables of the denoiser, ensuring that the estimated samples reside in a joint space of noise and conditions. The denoiser then directly learns the score in the corresponding conditional space. Subsequently, the difference between the scores generated under the conditional and unconditional labels is scaled using classifier-free guidance, yielding a final optimization direction, i.e., the predicted noise. Different samplers\cite{song2020score,song2020denoising,lu2022dpm} can then be introduced for generation.

However, directly applying the above training process to the speaker-listener scenario proposed in this paper presents certain challenges. First, concatenating the speaker and listener as a single variable for training with the diffusion model is inefficient. For the speaker and listener in the same interaction period within the distribution, their motion sequences originate from two distinct individuals, making it difficult to represent their independent roles through positional encoding. Moreover, the unsegmented input leads to feature interference between the roles, potentially slowing down the convergence of the training process. Additionally, when the same condition is applied to the latent variables of both the speaker and listener, the model struggles to distinguish between the roles, hindering effective training of the diffusion model in the speaker-listener interaction scenario. To address these issues, our proposed Inter-Diffusion optimizes the conditioning process of the denoiser to guide efficient noise prediction.

The training process is illustrated in Figure~\ref{fig:pipe}. First, we sample the speaker's motion \( x_0^s \) and the listener's motion \( x_0^l \) from the data distribution over a period of time. Note that the roles of speaker and listener switch as the speaking turn changes. Simultaneously, we sample a timestep \( t \) and gaussian noise \( \epsilon \), and obtain their noisy representations \( x_t^s \) and \( x_t^l \) via \( q(x_t | x_0) \). These noisy representations are then independently fed into the shallow modules of the denoiser and processed separately through the speaker flow and listener flow. To facilitate the model's understanding of the role attributes of each noisy variable for better conditional generation, we scale the speech conditions of the speaker and listener at the same timestep. Since the speech is produced by the speaker, this condition is strongly associated with the speaker and is assigned a higher weight. Conversely, the listener exhibits greater independence during listening, and a lower condition weight grants the listener more freedom to fit more realistic responses. Finally, we optimize the Inter-Diffusion denoiser using the variational lower bound loss of diffusion models.

By decoupling the speaker-listener interaction in this manner, the model can effortlessly learn the full-body motion representations of both the speaker and listener, enabling the simulation of more lifelike real-world interaction scenarios. In the following sections, we will provide a detailed introduction to our design specifics and processing modules.

\subsection{Diffusion Model in Gestures}

The commonly used conditional diffusion model reconstructs the original input from a set of noisy samples at different time steps controlled by a given variance sequence $\{\beta_{t}\}^{T}_{t=1}$. The training process consists of a forward process and a backward process. The noise distribution of the forward noising process at time step $t$ is defined as:

\begin{align}
\label{ddpm_q}
q\left(x_{t} \mid x_0\right) = \mathcal{N}\left(x_{t} ; \sqrt{1-\beta_{t}} x_0, \beta_{t} \mathbf{I}\right) \nonumber,
\end{align}

Subsequently, a parameterized backward process $p_{\theta}(x_{t-1}\mid x_t, c)$ is introduced to approximate the true posterior distribution $q(x_{t-1}\mid x_t, x_0)$ for denoising. The optimization objective is to maximize the following ELBO:
\begin{align}
    \underset{\theta}{\min} \mathop{\mathbb{E}}\limits_{t,\epsilon}\left[\infdiv{q(x_{t-1}\mid x_t, x_0)}{p_{{\theta}}(x_{t-1}\mid x_t,C)}\right].
\end{align}
where $C$ is the given control signal. In this process, $p_{\theta}(x_{t-1}\mid x_t, C)$ can be viewed as predicting a noise-free sample $\hat{x}_0$ and minimizing the $L_2$-loss between $\hat{x}_0$ and the data sample $x_0$, followed by sampling from the posterior distribution using $\hat{x}_0$.

In the task of human motion prediction, to better express realistic 3D human motions and enhance the physical properties of generated motions, additional constraints on different body parts can be introduced to improve the naturalness and coherence of the model output. For the generation process $p_{\theta}\left(x_{0: T}\right):=p\left(x_{T}\right) \prod_{t=1}^{T} p_{\theta}\left(x_{t-1} \mid x_{t}, C\right)$, starting from $x_T \sim \mathcal{N}(0,\mathbf{I})$, since the conditional diffusion model is obtained during training, the classifier-free guidance setting is followed to utilize a linear combination of the conditional noise predicted by the model:
\begin{align}
    \epsilon( x_t , t , c) = \epsilon_{\theta}( x_t , t , c) + s \cdot (\epsilon_{\theta}( x_t , t , c) - \epsilon_{\theta}( x_t , t , \phi))
\end{align}
Here, $s$ is a scaling factor greater than 1 and $\phi$ represents unconditional input. Different samplers are then used to iterate over the noisy samples to obtain a clear motion trajectory.
\subsection{Inter-Diffusion Structure}

The unmodified diffusion model directly inputs the noise sample $x_t$ into the denoiser for denoising. However, this does not align with the multi-role attribute action generation in the speaker-listener interaction scenario. Therefore, we need to redesign the data flow process of the noise sample $x_t$ in the denoiser. First, we define the latent variable at time $t$:

\begin{align}
x_t = x_t^s \oplus x_t^l    
\end{align}

where $\oplus$ represents the concatenation operator, $x_t^s, x_t^l \in \mathbb{R}^{2 \times D} $, respectively represent the noise representations of the speaker and listener in the latent space, where $D$ represents the dimension of the action features over the total time sequence. We additionally design a role encoding $r$ to prompt the denoiser about the role corresponding to the latent variable. Finally, $ x_t \in \mathbb{R}^{2 \times (D+r)}$ represents the joint representation of the actions of both at time $t$. In the following equations, we omit the role encoding $r$ for simplicity.


\textbf{Dual-Branch Architecture for Interactive Motion Generation}
Considering the autonomy and independence of the speaker and listener in motion generation, to avoid mutual interference of different attribute representations during joint representation, we adopt a divide-and-conquer approach. This allows the shallow layers of the denoiser to focus on extracting the respective role attribute features and reduces the context length to balance computational efficiency. We introduce a dual-branch Cross Local Attention (CLA) \cite{yang2023diffusestylegesture} to enable the model to deeply represent the noise variables \(x_t^s\),\(x_t^l\) and get the representation of
\begin{align}
z_s = CLA(x_t^s \oplus C_{\text{speaker}}) , \\
z_l = CLA(x_t^l \oplus C_{\text{listener}}),
\end{align}
where \(C_{\text{speaker}}\) and \(C_{\text{listener}}\) are the audio conditions for the inputs of different roles. Note that we reuse the same module when processing the representations of the speaker and listener. Here, the role encoding \(r\) we designed plays a key role, explicitly used to distinguish the role attributes of the input, ensuring that the local attention module adapts to the extraction patterns of different role features.

Subsequently, we integrate the global deep representations \(z = z_s \oplus z_l\) and introduce a self-attention mechanism to capture the interaction between the speaker and listener. To enhance the information interaction between the two, we split the output of the self-attention module and use the Cross Attention (CA) mechanism to capture the mutual dependencies between the deep representations of the speaker and listener, obtaining:
\begin{align}
\hat{x}_0^s = CA(z_s \oplus C_{\text{speaker}}) , \\
\hat{x}_0^l = CA(z_l \oplus C_{\text{listener}})
\end{align}
Finally, we concatenate the prediction results of the speaker and listener to obtain the final predicted sample \(\hat{x}_0 = \hat{x}_0^s \oplus \hat{x}_0^l\).

\textbf{Interactive Role Condition Representation}
In interactive dialogue scenarios, the audio condition $C$ comes from the initiator of the speech, i.e., the speaker, while the listener, as the receiver, often has more freedom to decide their body movements. Therefore, the speech condition's ability to constrain the actions of the two is not consistent, and the same condition input may affect the rationality of the listener's generated actions. To address this, we configure different condition inputs for the speaker and listener:
\begin{align}
C_{\text{speaker}} = \lambda C ,  \\
C_{\text{listener}} = 1-\lambda C
\end{align}
where \(\lambda \in [0,1]\) is the weight coefficient controlling the strength of the condition. The role encodings of the two characters in the dialogue process, along with the different conditions, help the network modules perceive the identity attributes of the input variables. As the network depth increases, the condition signal weakens, and we re-concatenate the condition signal at various cross-variable information exchange stages to improve the guidance effect.

\textbf{Loss Function}
The efficient network design of Inter-Diffusion ensures that the denoiser can be trained without the need for complex loss functions. Our loss function consists of two terms. The first term is the variational lower bound loss \(\mathcal{L}_{VLB}\), which can be further simplified as:
\begin{align}
\mathcal{L}_{\text{simple}} = \mathbb{E}_{t, x_0, \epsilon} \left[ \| \epsilon - \epsilon_\theta(x_t, t, C) \|^2 \right]
\end{align}
Additionally, we introduce the geometric loss \(\mathcal{L}_{foot}\) from MDM to strengthen the physical constraints on the foot. Ultimately, our loss function is:
\begin{align}
\mathcal{L} = \mathcal{L}_{\text{simple}} + \alpha \mathcal{L}_{\text{foot}}
\end{align}

\section{Experiments}
In this section, we evaluate the performance of our interactive diffusion-based system for generating gestures of both speakers and listeners. We compare it with existing methods to highlight its superiority in capturing two-way communication. Ablation studies are conducted to validate the key modules in our system. Generalization experiments explore the method's potential in different virtual communication scenarios. Considering the complexity of human gestures in speaker - listener interactions, we carry out user studies. These studies help prove the system's high quality performance. 
\subsection{Dataset Design and Evaluation Metrics}
\textbf{Datasets.} In our experiment, we focus on processing data from the TWH and ZeroEGG datasets to train our model for generating gestures of both speakers and listeners. For the TWH dataset, we first isolate the listener data. Since the listener's motion data might not be directly compatible with the audio in the ZeroEGG dataset in terms of length and synchronization, we employ the slerp interpolation method. This allows us to reverse - connect the listener's actions in the TWH dataset, ensuring that the length of the listener's motion sequences can match the audio length in the ZeroEGG dataset. This processed data serves as the listener data during the training process.

We assign style labels to distinguish between speakers and listeners. Specifically, we use 0 as the style label for speakers and 1 for listeners. When switching between the generation of speaker and listener gestures, we adjust the weights of the audio features and gesture features. For instance, when generating speaker - related gestures, we assign a relatively higher weight to the audio features, as the speaker's gestures are more directly related to the content of the speech. Conversely, when generating listener - related gestures, we place more emphasis on the gesture features, considering that the listener's responses are more about non - verbal feedback. This data processing method not only enables us to make full use of the data from different datasets but also effectively trains the model to generate context - appropriate gestures for both speakers and listeners. We then use this processed data to train our inter - diffusion model, and later evaluate the model's performance on a separate test set to validate the effectiveness of our data - processing approach.

\noindent\textbf{Evaluation Metrics.} 
In order to comprehensively evaluate the performance of our proposed method in generating natural and context - appropriate gestures for both speakers and listeners, we adopt a set of well-defined evaluation indicators. These indicators cover crucial aspects such as gesture quality, alignment with audio beats, and diversity, providing a multi-faceted assessment of our model's capabilities. 

We measure gesture quality with the Fréchet Gesture Distance (FGD) \cite{yoon2020speech}. FGD calculates the distance between the latent feature distributions of generated and real - world gestures. A lower FGD value means the generated gestures are more like real life ones, indicating higher quality. Beat Alignment (BA) \cite{li2021ai} evaluates the synchronization between generated gestures and audio beats. It's calculated as the inverse angular distance between them. A higher BA value shows better alignment, enhancing the naturalness of the interaction. We use DIV to assess gesture diversity. DIV measures the L1 distance among gestures generated under the same control signal. A larger DIV value means the model can generate more diverse gestures, capturing the natural variability in human non-verbal communication.

\textbf{Implementation Details}. The Inter-Diffusion Generation Model is built upon a novel architecture designed to handle the complex interaction between speakers and listeners. At its core, the Inter-Diff mechanism consists of two main submodules: the speaker - centric diffusion module and the listener-response diffusion module.

The speaker-centric diffusion module takes the speaker's audio and semantic information as inputs. It uses a multilayer neural network with convolutional and recurrent layers to extract high level features. These features are then fed into the diffusion process, which gradually denoises the initial noise vector to generate the speaker's gesture sequences.

The listener-response diffusion module, on the other hand, receives the output from the speaker centric module as well as the listener's own context information. It is designed to generate appropriate listener gestures in response to the speaker. This module also employs a combination of convolutional and recurrent layers to capture the relevant information and generate context-aware listener gestures. The two modules interact with each other through a set of cross - module connections, allowing the model to capture the dynamic nature of the speaker-listener interaction.

We trained the model using a single A100 GPU. The training process lasted for 72 hours with a batch size of 96. For the optimizer, we used Adam. The learning rate for the generator was set to 2e - 5, and for the discriminator, it was 1e - 4. The betas for the Adam optimizer were set to (0.9, 0.999), and the weight decay was set to 1e - 5. During the training, we monitored the loss functions of both the generator and the discriminator to ensure stable convergence.

\subsection{Comparison with Contemporary Methods}
In this experiment, we aim to rigorously assess the generation quality of our method in contrast to contemporary techniques, with a particular focus on leveraging the unique characteristics of the ZeroEGG and TWH datasets. For the ZeroEGG dataset, we explore a wide array of styles including happiness, sadness, anger, aging, neutral, fatigue, and relaxation. This diverse set of styles enables a comprehensive evaluation of our method's ability to generate context - appropriate gestures across various emotional and situational contexts. From the TWH dataset, we extract the listener - related data, which plays a crucial role in evaluating the performance of our model in generating natural listener gestures.

All compared methods, like ours, utilize WavLM for audio feature extraction to ensure a fair comparison ground. We partition both the ZeroEGG and TWH datasets into training, validation, and test sets with proportions of 0.8, 0.1, and 0.1 respectively. Subsequently, we train all models on the entire training set.

When it comes to evaluating the generation quality, we adopt different strategies for speakers and listeners. For speaker - related gesture generation, we use objective metrics such as Fréchet Gesture Distance (FGD). FGD calculates the distance between the latent feature distributions of generated and real - world gestures, providing a quantifiable measure of how closely the generated speaker gestures mimic real - life ones. However, due to the non - standard nature of the communication data related to listeners, objective metrics are not well - suited to evaluate the effectiveness of listener gesture generation. Instead, we conduct a user study for subjective evaluation. In this user study, we recruit a diverse group of participants. They are presented with the generated listener gestures along with the corresponding audio and context. The participants are then asked to rate the naturalness, context - appropriateness, and overall quality of the listener gestures on a predefined scale. This subjective evaluation approach allows us to capture the nuances and subtleties in listener non - verbal communication that objective metrics might miss.

\begin{table}
\resizebox{1.0\linewidth}{!}{
\begin{tabular}{lcccc}
\toprule[1.5pt]
\multirow{2}{*}{Method} & \multicolumn{3}{c}{ZeroEGG} & \\ \cline{2-4}
& FGD$\downarrow$   & BA$\uparrow$ & DIV$\uparrow$  &  \\ \cline{2-4} \hline
DSG~\cite{yang2023diffusestylegesture}            & 15.67     & 0.81 &0.63     \\ 
FreeTalker~\cite{yang2024freetalker}    &17.12 &0.72 &\textbf{0.84} \\
DiffGesture(re-train)~\cite{zhu2023taming}  &25.7 &0.58 &- \\
\textbf{Ours}           & \textbf{14.57}    &\textbf{0.82}  &  0.83   \\
\hline
Trimodal(re-train)~\cite{yoon2022genea} &22.4 &0.72 &0.80 \\
HA2G(re-train)~\cite{liu2022audio}&18.8 &0.81 &0.73 \\
CAMN~\cite{liu2022beat} &15.52 &0.83 &0.77   \\ \hline
\end{tabular}
}
\caption{Objective Metrics. The FGD evaluation model is trained across the entire training set, with lower evaluation values indicating closer adherence to the original motion distribution. }
\vspace{-15pt}
\label{tab:obj}
\end{table}

\begin{figure}
  \begin{center}
    \includegraphics[width=1\linewidth]{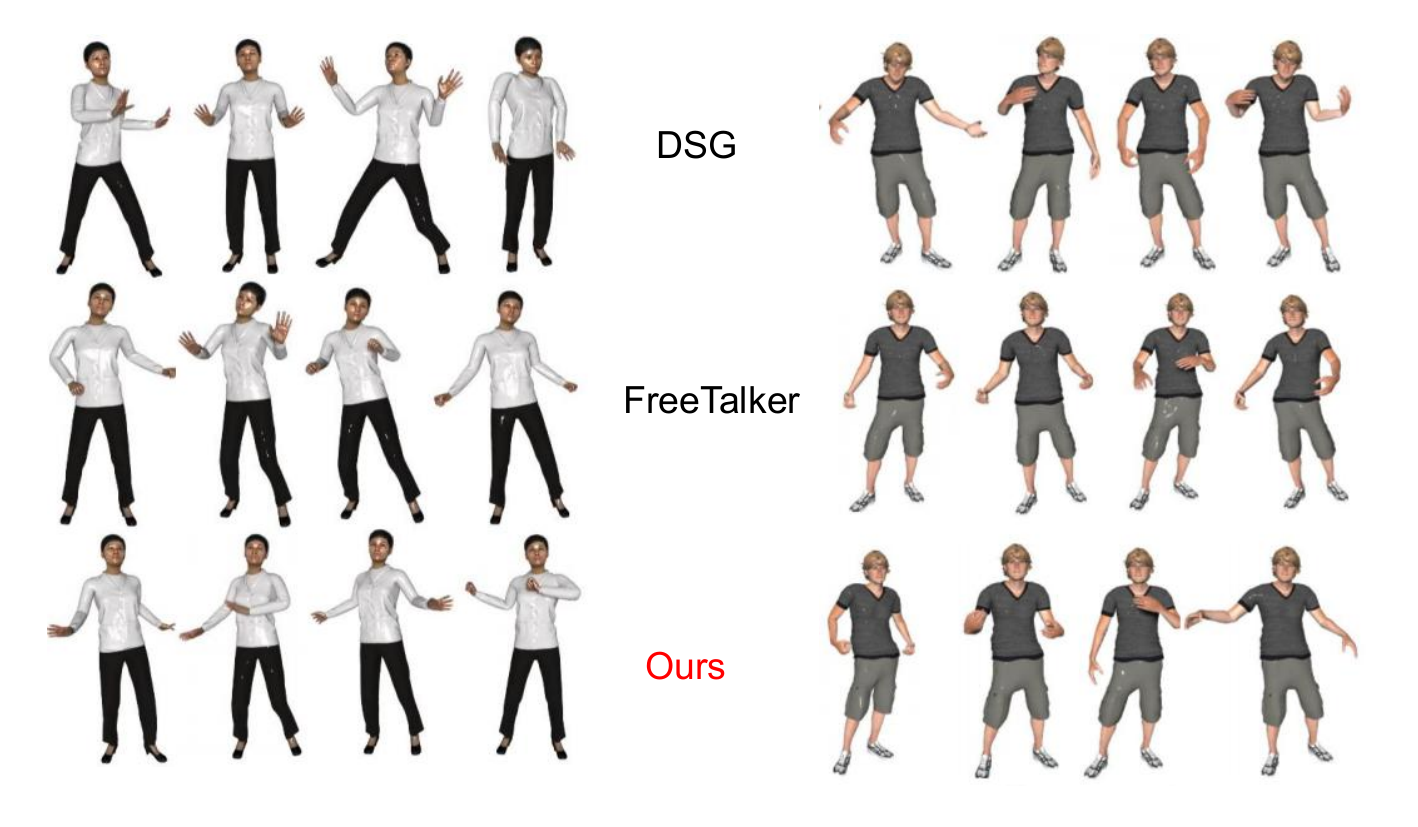}
  \end{center}
  \vspace{-10pt}
  \caption{In comparison to contemporary diffusion-based methods such as DSG and FreeTalker, our approach achieves the best result.}
  \Description[<short description>]{}
  \label{fig:qual}
\end{figure}

As shown in the Table~\ref{tab:obj} and Figure~\ref{fig:qual}, our method demonstrates remarkable performance in generating gestures on the ZeroEGG dataset. In the FGD metric, which gauges the similarity between generated and real - world gestures, our method scores 14.57. This is lower than DSG's~\cite{yang2023diffusestylegesture} 15.67, FreeTalker's~\cite{yang2024freetalker} 17.12, DiffGesture's~\cite{zhu2023taming} 25.7, Trimodal's~\cite{yoon2022genea} 22.4~\cite{liu2022audio}, and HA2G's~\cite{liu2022beat} 18.8, indicating that our generated gestures are more closely aligned with the original motion distribution. Regarding BA, which measures the synchronization between audio and gesture beats, our method attains a value of 0.82. This figure is higher than most of the compared methods, such as DSG's 0.81, FreeTalker's 0.72, DiffGesture's 0.58, Trimodal's 0.72, and HA2G's 0.81, second only to CAMN's 0.83. This result showcases the strong audio - gesture synchronization of our generated gestures. In terms of DIV, which reflects the diversity of gestures, although FreeTalker leads with 0.84, our method closely follows with 0.83, outperforming DiffGesture (not available), Trimodal's 0.80, HA2G's 0.73, and CAMN's 0.77. This indicates that our model can generate a diverse range of gestures.

\begin{table*}

\centering
\resizebox{1.0\linewidth}{!}{
\begin{tabular}{lcccc} 
\toprule[1.5pt]
\multirow{1}{*}{Method} & \multicolumn{1}{c}{\textit{Human-likeness}} & \multicolumn{1}{c}{\textit{Gesture-appropriateness}} & \multicolumn{1}{c}{\textit{Emotion-compatibility}} &\multicolumn{1}{c}{\textit{Listener-coherence}}\\ \hline
Ground Truth & $4.61 \pm 0.17$ & $4.72 \pm 0.20$ &$3.62 \pm 0.27$ & $4.12 \pm 0.31$\\ 
DSG~\cite{yang2023diffusestylegesture} & $3.46 \pm 0.16$ & $3.61 \pm 0.11$ & $3.67 \pm 0.19$ & $3.01 \pm 0.12$ \\ 
DiffGesture~\cite{zhu2023taming} & $2.92 \pm 0.22$ & $3.87 \pm 0.09$ & $1.22 \pm 0.19$ & $3.01 \pm 0.12$ \\\hline
\textbf{Ours} & $4.17 \pm 0.13$ &$4.20 \pm 0.18$ &$3.88 \pm 0.21$ & $4.01 \pm 0.17$ \\

\hline
\end{tabular}
}
\caption{\textbf{95\% Confidence Interval for User Study Average Score.} The metrics include "Human - likeness", which measures how similar the generated gestures are to natural human gestures. "Gesture - appropriateness" assesses whether the gestures are suitable in a given context. "Emotion - compatibility" evaluates the consistency between the emotions expressed by the gestures and the associated audio or text. The table compares Ground Truth, DiffStyleGesture, and our method (Ours). The scores, with the values after "±" representing the 95\% confidence interval, show the relative performance of each method in these subjective aspects.
}
\label{tab:user}
\end{table*}

\subsection{User study} To thoroughly evaluate the real - world visual performance of our method, we carried out a well - designed user study. In this study, we made direct comparisons with DSG, DiffGesture, and the ground truth data, which served as crucial benchmarks.
We started by carefully curating 12 audio clips from the test set. These clips were selected to ensure a wide variety in terms of speaking styles, emotional tones, and topics. Subsequently, we segmented them into short and long segments. The short segments were used to test the model's ability to generate appropriate gestures for quick - paced, concise expressions, while the long segments aimed to evaluate how well the model could handle more complex, extended conversations.
A total of 30 participants, with diverse backgrounds including those from the fields of computer - human interaction, multimedia design, and general users interested in virtual communication, were invited to take part. They were presented with the gesture sequences that were meticulously rendered as virtual speakers within the Blender environment. This high - fidelity rendering allowed the participants to focus solely on the quality and naturalness of the gestures.
We employed four key metrics, each on a 1 - 5 scale, to objectively measure the performance. "Human - likeness" was used to gauge how closely the generated gestures resembled those of real humans. A higher score here indicates that the model - generated movements were more in line with natural human motion patterns. "Gesture - appropriateness" evaluated whether the gestures were suitable within the given context. For example, a thumbs - up gesture in a positive - feedback context would score higher. "Emotion - compatibility" focused on the alignment between the emotions expressed through the gestures and the audio content. If the audio conveyed happiness and the gestures mirrored that emotion, a higher score was given. "Listener - coherence" examined the smoothness and logical connection of the listener's gestures during the communication process, assessing how well they interacted with the speaker's actions.
As clearly shown in Table~\ref{tab:user}, our method demonstrated significant superiority over DSG and DiffGesture in most of these metrics. For "Human - likeness", we achieved a score of \(4.17 \pm 0.13\), which is notably higher than the other two methods, indicating that our generated gestures are more authentic in terms of human movement. In "Gesture - appropriateness", our score of \(4.20 \pm 0.18\) shows that our model is better at generating context - relevant gestures. Regarding "Emotion - compatibility", our score of \(3.88 \pm 0.21\) reveals that the emotions expressed by our generated gestures are more in tune with the audio. And for "Listener - coherence", with a score of \(4.01 \pm 0.17\), it is evident that our model can generate more natural - flowing listener gestures.

\subsection{Ablation Study}
\begin{table}

\centering
\resizebox{1.0\linewidth}{!}{
\begin{tabular}{lccccccccc}
\toprule[1.5pt]
\multirow{1}{*}{Method} & \multicolumn{1}{c}{\textit{$FGD$}} & \multicolumn{1}{c}{\textit{BA}} & \multicolumn{1}{c}{\textit{DIV}}\\ \hline
DSG~\cite{yang2023diffusestylegesture} & 25.7 & 0.81 &0.63 \\ \hline
Baseline  & 25.9 & 0.70 &0.57  \\
Baseline w/ Inter-Diff  & 20.1 & 0.63 & 0.77\\
Baseline w/ cross-local attention  & 18.3 & 0.82 &0.64  \\
Baseline w/ state-weight  & 16.9 & 0.75 &0.71  \\
Ours  & 14.57 & 0.82 &0.83    \\ 
\hline
\end{tabular}
}
\caption{\textbf{Ablation study.} Shows the effect of different modules on the model.} 
\label{tab:abl}
\end{table}
As shown in Table~\ref{tab:abl}, we meticulously carried out an ablation study with the aim of comprehensively exploring the intricate impact of different modules on the model's performance. Our baseline model, which was constructed based on MDM~\cite{tevet2022human}, initially served as a crucial reference point. It exhibited an FGD (Fréchet Gesture Distance) of 25.9, a BA (Beat Alignment) of 0.70, and a DIV (Diversity) of 0.57. These baseline metrics provided a benchmark against which the contributions of additional modules could be precisely measured.

When we introduced the Inter - Diff module to the baseline, resulting in the "Baseline w/ Inter - Diff" configuration, several significant changes occurred. The FGD, which measures the similarity between the latent feature distributions of generated and real - world gestures, dropped substantially from 25.9 to 20.1. This decrease clearly indicates that the Inter - Diff module plays a pivotal role in enhancing the naturalness of the generated gestures. A lower FGD implies that the model - generated gestures are more closely aligned with the real - life motion patterns, making them appear more authentic. Simultaneously, the DIV increased from 0.57 to 0.77. This upward shift in the diversity metric showcases the Inter - Diff module's ability to endow the model with the capacity to generate a wider variety of gestures. However, it's worth noting that the BA decreased slightly from 0.70 to 0.63. This reduction might be attributed to the fact that the Inter - Diff module's primary focus on enhancing naturalness and diversity could have, to some extent, affected the strict alignment of gestures with audio beats. But overall, the positive impacts on naturalness and diversity are substantial.

For the "Baseline w/ cross - local attention" model, the FGD reached 18.3, which is a significant improvement compared to the baseline. This indicates that the cross - local attention mechanism effectively enhances the quality of the generated gestures. A lower FGD value suggests that the model is better at capturing the fine - grained details of real - world gestures. Moreover, the BA hit 0.82, which is the highest among some of the ablated models. This high BA value demonstrates that the cross - local attention module is highly effective in aligning the generated gestures with the audio beats, making the interaction between the audio and gestures more synchronized. However, the DIV of 0.64 indicates that while the cross - local attention module excels in quality and alignment, the diversity of the generated gestures is only moderate. It might be that the focus on alignment and quality comes at the expense of generating a highly diverse set of gestures.

The "Baseline w/ state - weight" model achieved an FGD of 16.9 and a BA of 0.75. The relatively low FGD value shows that the state - weight module contributes positively to the generation of high - quality gestures, bringing the generated gestures closer to the real - world motion distribution. The BA of 0.75 also indicates a reasonable level of alignment between the gestures and the audio. Along with a DIV of 0.71, it becomes evident that the state - weight module has a positive impact on both the quality and diversity of gesture generation. It strikes a balance between generating natural - looking gestures and maintaining a certain level of diversity.

Our complete method, which incorporates all the essential modules, outperformed all the ablated versions. With an FGD of 14.57, it achieved the lowest FGD value among all, indicating that our full - fledged model is the best at generating gestures that closely resemble real - world ones. The BA of 0.82 is on par with the highest - achieving ablated model in terms of beat alignment, ensuring that the gestures are well - synchronized with the audio. And the DIV of 0.83 represents the highest diversity among all the models. This combination of the lowest FGD, high BA, and high DIV clearly demonstrates the combined strength of all the modules in our model. It validates the importance of each module and their synergy in creating a high - performing model that can generate high - quality, well - aligned, and diverse gestures.

\section{Applications}
Our application, which is uniquely powered by the method of incorporating listener-based gesture generation, represents a significant leap forward in the realm of virtual interaction. This innovative approach has enabled us to create a virtual communication environment that is far more natural and immersive than what was previously possible.

In the applications, the integration of listener-based gesture generation completely transforms the way users interact with virtual entities. When a user initiates communication, the virtual entities are not only capable of providing verbal responses but also generate a rich array of non-verbal cues. The generation of listening gestures, such as appropriate head movements, body postures, and hand gestures that signal attentiveness, creates a sense of genuine engagement. These gestures are not only context-sensitive but also adapt dynamically based on the flow of the conversation.



\section{Limitations}
Despite the promising results, our approach has several limitations. First, we still rely on the traditional Markov - chain - based DDPM strategy. This strategy, although effective~\cite{cheng2025didiffges, cheng2025conditional, cheng2025hologest} in generating high - quality gestures, poses a significant drawback in terms of generation speed. As a result, our model may not be suitable for some potential real - time applications that demand rapid gesture generation. For instance, in real - time virtual meetings or live gaming scenarios where immediate response is crucial, the relatively slow generation speed can lead to a delay in the interaction, reducing the overall user experience.

Moreover, the current model is highly dependent on the quality and quantity of the training data. The scarcity of comprehensive listener - related data, especially in some rare communication scenarios, restricts the model's ability to generalize well. This may result in less - accurate gesture generation when dealing with less - common communication contexts.

\section{Conclusion}
In this research, we introduced the Inter-Diffusion Generation Model, a novel approach that revolutionizes the landscape of gesture generation in virtual communication. At the core of our work is the introduction of an interactive diffusion mechanism, which enables the generation of both speaker and listener gestures. This model captures the dynamic interplay between speakers and listeners. The gestures of speakers influence the listener's non-verbal responses, and vice versa. By training on the combined TWH and zeroegg datasets, our model learned to generate context - appropriate and natural - looking gestures for both parties. This research paves the way for more natural virtual communication experiences. It has potential applications in various fields such as virtual reality, gaming, and film, opening up new possibilities for creating immersive and interactive virtual environments. Considering the expansion of the dataset in the future, we will also introduce some motion capture methods~\cite{cheng2023bopr, liang2025ropetp} to improve the quality of the dataset.


\bibliographystyle{ACM-Reference-Format}
\bibliography{sample-base}


\begin{thebibliography}{32}


\ifx \showCODEN    \undefined \def \showCODEN     #1{\unskip}     \fi
\ifx \showISBNx    \undefined \def \showISBNx     #1{\unskip}     \fi
\ifx \showISBNxiii \undefined \def \showISBNxiii  #1{\unskip}     \fi
\ifx \showISSN     \undefined \def \showISSN      #1{\unskip}     \fi
\ifx \showLCCN     \undefined \def \showLCCN      #1{\unskip}     \fi
\ifx \shownote     \undefined \def \shownote      #1{#1}          \fi
\ifx \showarticletitle \undefined \def \showarticletitle #1{#1}   \fi
\ifx \showURL      \undefined \def \showURL       {\relax}        \fi
\providecommand\bibfield[2]{#2}
\providecommand\bibinfo[2]{#2}
\providecommand\natexlab[1]{#1}
\providecommand\showeprint[2][]{arXiv:#2}

\bibitem[Ahuja et~al\mbox{.}(2020)]%
        {ahuja2020style}
\bibfield{author}{\bibinfo{person}{Chaitanya Ahuja}, \bibinfo{person}{Dong~Won Lee}, \bibinfo{person}{Yukiko~I Nakano}, {and} \bibinfo{person}{Louis-Philippe Morency}.} \bibinfo{year}{2020}\natexlab{}.
\newblock \showarticletitle{Style transfer for co-speech gesture animation: A multi-speaker conditional-mixture approach}. In \bibinfo{booktitle}{\emph{Computer Vision--ECCV 2020: 16th European Conference, Glasgow, UK, August 23--28, 2020, Proceedings, Part XVIII 16}}. Springer, \bibinfo{pages}{248--265}.
\newblock


\bibitem[Ao et~al\mbox{.}(2023)]%
        {ao2023gesturediffuclip}
\bibfield{author}{\bibinfo{person}{Tenglong Ao}, \bibinfo{person}{Zeyi Zhang}, {and} \bibinfo{person}{Libin Liu}.} \bibinfo{year}{2023}\natexlab{}.
\newblock \showarticletitle{GestureDiffuCLIP: Gesture Diffusion Model with {CLIP} Latents}.
\newblock \bibinfo{journal}{\emph{{ACM} Trans. Graph.}} \bibinfo{volume}{42}, \bibinfo{number}{4} (\bibinfo{year}{2023}), \bibinfo{pages}{42:1--42:18}.
\newblock
\href{https://doi.org/10.1145/3592097}{doi:\nolinkurl{10.1145/3592097}}


\bibitem[Cheng and Huang(2025)]%
        {cheng2025hologest}
\bibfield{author}{\bibinfo{person}{Yongkang Cheng} {and} \bibinfo{person}{Shaoli Huang}.} \bibinfo{year}{2025}\natexlab{}.
\newblock \showarticletitle{HoloGest: Decoupled Diffusion and Motion Priors for Generating Holisticly Expressive Co-speech Gestures}.
\newblock \bibinfo{journal}{\emph{arXiv preprint arXiv:2503.13229}} (\bibinfo{year}{2025}).
\newblock


\bibitem[Cheng et~al\mbox{.}(2025a)]%
        {cheng2025didiffges}
\bibfield{author}{\bibinfo{person}{Yongkang Cheng}, \bibinfo{person}{Shaoli Huang}, \bibinfo{person}{Xuelin Chen}, \bibinfo{person}{Jifeng Ning}, {and} \bibinfo{person}{Mingming Gong}.} \bibinfo{year}{2025}\natexlab{a}.
\newblock \showarticletitle{DIDiffGes: Decoupled Semi-Implicit Diffusion Models for Real-time Gesture Generation from Speech}. In \bibinfo{booktitle}{\emph{Proceedings of the AAAI Conference on Artificial Intelligence}}, Vol.~\bibinfo{volume}{39}. \bibinfo{pages}{2464--2472}.
\newblock


\bibitem[Cheng et~al\mbox{.}(2023)]%
        {cheng2023bopr}
\bibfield{author}{\bibinfo{person}{Yongkang Cheng}, \bibinfo{person}{Shaoli Huang}, \bibinfo{person}{Jifeng Ning}, {and} \bibinfo{person}{Ying Shan}.} \bibinfo{year}{2023}\natexlab{}.
\newblock \showarticletitle{Bopr: Body-aware part regressor for human shape and pose estimation}.
\newblock \bibinfo{journal}{\emph{arXiv preprint arXiv:2303.11675}} (\bibinfo{year}{2023}).
\newblock


\bibitem[Cheng et~al\mbox{.}(2025b)]%
        {cheng2025conditional}
\bibfield{author}{\bibinfo{person}{Yongkang Cheng}, \bibinfo{person}{Mingjiang Liang}, \bibinfo{person}{Shaoli Huang}, \bibinfo{person}{Gaoge Han}, \bibinfo{person}{Jifeng Ning}, {and} \bibinfo{person}{Wei Liu}.} \bibinfo{year}{2025}\natexlab{b}.
\newblock \showarticletitle{Conditional gan for enhancing diffusion models in efficient and authentic global gesture generation from audios}. In \bibinfo{booktitle}{\emph{2025 IEEE/CVF Winter Conference on Applications of Computer Vision (WACV)}}. IEEE, \bibinfo{pages}{2164--2173}.
\newblock


\bibitem[Cheng et~al\mbox{.}(2024)]%
        {cheng2024expgest}
\bibfield{author}{\bibinfo{person}{Yongkang Cheng}, \bibinfo{person}{Mingjiang Liang}, \bibinfo{person}{Shaoli Huang}, \bibinfo{person}{Jifeng Ning}, {and} \bibinfo{person}{Wei Liu}.} \bibinfo{year}{2024}\natexlab{}.
\newblock \showarticletitle{Expgest: Expressive speaker generation using diffusion model and hybrid audio-text guidance}. In \bibinfo{booktitle}{\emph{2024 IEEE International Conference on Multimedia and Expo (ICME)}}. IEEE, \bibinfo{pages}{1--6}.
\newblock


\bibitem[Habibie et~al\mbox{.}(2022)]%
        {habibie2022motion}
\bibfield{author}{\bibinfo{person}{Ikhsanul Habibie}, \bibinfo{person}{Mohamed Elgharib}, \bibinfo{person}{Kripasindhu Sarkar}, \bibinfo{person}{Ahsan Abdullah}, \bibinfo{person}{Simbarashe Nyatsanga}, \bibinfo{person}{Michael Neff}, {and} \bibinfo{person}{Christian Theobalt}.} \bibinfo{year}{2022}\natexlab{}.
\newblock \showarticletitle{A Motion Matching-based Framework for Controllable Gesture Synthesis from Speech}. In \bibinfo{booktitle}{\emph{ACM SIGGRAPH 2022 Conference Proceedings}}. \bibinfo{pages}{1--9}.
\newblock


\bibitem[Habibie et~al\mbox{.}(2021)]%
        {habibie2021learning}
\bibfield{author}{\bibinfo{person}{Ikhsanul Habibie}, \bibinfo{person}{Weipeng Xu}, \bibinfo{person}{Dushyant Mehta}, \bibinfo{person}{Lingjie Liu}, \bibinfo{person}{Hans-Peter Seidel}, \bibinfo{person}{Gerard Pons-Moll}, \bibinfo{person}{Mohamed Elgharib}, {and} \bibinfo{person}{Christian Theobalt}.} \bibinfo{year}{2021}\natexlab{}.
\newblock \showarticletitle{Learning speech-driven 3d conversational gestures from video}. In \bibinfo{booktitle}{\emph{Proceedings of the 21st ACM International Conference on Intelligent Virtual Agents}}. \bibinfo{pages}{101--108}.
\newblock


\bibitem[Han et~al\mbox{.}(2025)]%
        {han2025reindiffuse}
\bibfield{author}{\bibinfo{person}{Gaoge Han}, \bibinfo{person}{Mingjiang Liang}, \bibinfo{person}{Jinglei Tang}, \bibinfo{person}{Yongkang Cheng}, \bibinfo{person}{Wei Liu}, {and} \bibinfo{person}{Shaoli Huang}.} \bibinfo{year}{2025}\natexlab{}.
\newblock \showarticletitle{Reindiffuse: Crafting physically plausible motions with reinforced diffusion model}. In \bibinfo{booktitle}{\emph{2025 IEEE/CVF Winter Conference on Applications of Computer Vision (WACV)}}. IEEE, \bibinfo{pages}{2218--2227}.
\newblock


\bibitem[Ho et~al\mbox{.}(2020)]%
        {ho2020denoising}
\bibfield{author}{\bibinfo{person}{Jonathan Ho}, \bibinfo{person}{Ajay Jain}, {and} \bibinfo{person}{Pieter Abbeel}.} \bibinfo{year}{2020}\natexlab{}.
\newblock \showarticletitle{Denoising diffusion probabilistic models}.
\newblock \bibinfo{journal}{\emph{Advances in neural information processing systems}}  \bibinfo{volume}{33}, \bibinfo{pages}{6840--6851}.
\newblock


\bibitem[Ho and Salimans(2022)]%
        {ho2022classifier}
\bibfield{author}{\bibinfo{person}{Jonathan Ho} {and} \bibinfo{person}{Tim Salimans}.} \bibinfo{year}{2022}\natexlab{}.
\newblock \showarticletitle{Classifier-free diffusion guidance}.
\newblock \bibinfo{journal}{\emph{arXiv preprint arXiv:2207.12598}} (\bibinfo{year}{2022}).
\newblock


\bibitem[Li et~al\mbox{.}(2021)]%
        {li2021ai}
\bibfield{author}{\bibinfo{person}{Ruilong Li}, \bibinfo{person}{Shan Yang}, \bibinfo{person}{David~A Ross}, {and} \bibinfo{person}{Angjoo Kanazawa}.} \bibinfo{year}{2021}\natexlab{}.
\newblock \showarticletitle{Ai choreographer: Music conditioned 3d dance generation with aist++}. In \bibinfo{booktitle}{\emph{Proceedings of the IEEE/CVF International Conference on Computer Vision}}. \bibinfo{pages}{13401--13412}.
\newblock


\bibitem[Liang et~al\mbox{.}(2025)]%
        {liang2025ropetp}
\bibfield{author}{\bibinfo{person}{Mingjiang Liang}, \bibinfo{person}{Yongkang Cheng}, \bibinfo{person}{Hualin Liang}, \bibinfo{person}{Shaoli Huang}, {and} \bibinfo{person}{Wei Liu}.} \bibinfo{year}{2025}\natexlab{}.
\newblock \showarticletitle{Ropetp: Global human motion recovery via integrating robust pose estimation with diffusion trajectory prior}. In \bibinfo{booktitle}{\emph{2025 IEEE/CVF Winter Conference on Applications of Computer Vision (WACV)}}. IEEE, \bibinfo{pages}{2973--2982}.
\newblock


\bibitem[Liu et~al\mbox{.}(2022b)]%
        {liu2022beat}
\bibfield{author}{\bibinfo{person}{Haiyang Liu}, \bibinfo{person}{Zihao Zhu}, \bibinfo{person}{Naoya Iwamoto}, \bibinfo{person}{Yichen Peng}, \bibinfo{person}{Zhengqing Li}, \bibinfo{person}{You Zhou}, \bibinfo{person}{Elif Bozkurt}, {and} \bibinfo{person}{Bo Zheng}.} \bibinfo{year}{2022}\natexlab{b}.
\newblock \showarticletitle{Beat: A large-scale semantic and emotional multi-modal dataset for conversational gestures synthesis}. In \bibinfo{booktitle}{\emph{European Conference on Computer Vision}}. Springer, \bibinfo{pages}{612--630}.
\newblock


\bibitem[Liu et~al\mbox{.}(2022a)]%
        {liu2022audio}
\bibfield{author}{\bibinfo{person}{Xian Liu}, \bibinfo{person}{Qianyi Wu}, \bibinfo{person}{Hang Zhou}, \bibinfo{person}{Yuanqi Du}, \bibinfo{person}{Wayne Wu}, \bibinfo{person}{Dahua Lin}, {and} \bibinfo{person}{Ziwei Liu}.} \bibinfo{year}{2022}\natexlab{a}.
\newblock \showarticletitle{Audio-Driven Co-Speech Gesture Video Generation}.
\newblock \bibinfo{journal}{\emph{Advances in Neural Information Processing Systems}}  \bibinfo{volume}{35}, \bibinfo{pages}{21386--21399}.
\newblock


\bibitem[Lu et~al\mbox{.}(2022)]%
        {lu2022dpm}
\bibfield{author}{\bibinfo{person}{Cheng Lu}, \bibinfo{person}{Yuhao Zhou}, \bibinfo{person}{Fan Bao}, \bibinfo{person}{Jianfei Chen}, \bibinfo{person}{Chongxuan Li}, {and} \bibinfo{person}{Jun Zhu}.} \bibinfo{year}{2022}\natexlab{}.
\newblock \showarticletitle{Dpm-solver: A fast ode solver for diffusion probabilistic model sampling in around 10 steps}.
\newblock \bibinfo{journal}{\emph{Advances in Neural Information Processing Systems}}  \bibinfo{volume}{35} (\bibinfo{year}{2022}), \bibinfo{pages}{5775--5787}.
\newblock


\bibitem[Song et~al\mbox{.}(2020a)]%
        {song2020denoising}
\bibfield{author}{\bibinfo{person}{Jiaming Song}, \bibinfo{person}{Chenlin Meng}, {and} \bibinfo{person}{Stefano Ermon}.} \bibinfo{year}{2020}\natexlab{a}.
\newblock \showarticletitle{Denoising diffusion implicit models}.
\newblock \bibinfo{journal}{\emph{arXiv preprint arXiv:2010.02502}} (\bibinfo{year}{2020}).
\newblock


\bibitem[Song et~al\mbox{.}(2020b)]%
        {song2020score}
\bibfield{author}{\bibinfo{person}{Yang Song}, \bibinfo{person}{Jascha Sohl-Dickstein}, \bibinfo{person}{Diederik~P Kingma}, \bibinfo{person}{Abhishek Kumar}, \bibinfo{person}{Stefano Ermon}, {and} \bibinfo{person}{Ben Poole}.} \bibinfo{year}{2020}\natexlab{b}.
\newblock \showarticletitle{Score-based generative modeling through stochastic differential equations}.
\newblock \bibinfo{journal}{\emph{arXiv preprint arXiv:2011.13456}} (\bibinfo{year}{2020}).
\newblock


\bibitem[Tevet et~al\mbox{.}(2022)]%
        {tevet2022human}
\bibfield{author}{\bibinfo{person}{Guy Tevet}, \bibinfo{person}{Sigal Raab}, \bibinfo{person}{Brian Gordon}, \bibinfo{person}{Yonatan Shafir}, \bibinfo{person}{Daniel Cohen-Or}, {and} \bibinfo{person}{Amit~H Bermano}.} \bibinfo{year}{2022}\natexlab{}.
\newblock \showarticletitle{Human motion diffusion model}.
\newblock \bibinfo{journal}{\emph{arXiv preprint arXiv:2209.14916}}.
\newblock


\bibitem[Wang et~al\mbox{.}(2022)]%
        {wang2022diffusion}
\bibfield{author}{\bibinfo{person}{Zhendong Wang}, \bibinfo{person}{Huangjie Zheng}, \bibinfo{person}{Pengcheng He}, \bibinfo{person}{Weizhu Chen}, {and} \bibinfo{person}{Mingyuan Zhou}.} \bibinfo{year}{2022}\natexlab{}.
\newblock \showarticletitle{Diffusion-gan: Training gans with diffusion}.
\newblock \bibinfo{journal}{\emph{arXiv preprint arXiv:2206.02262}} (\bibinfo{year}{2022}).
\newblock


\bibitem[Xie et~al\mbox{.}(2022)]%
        {xie2022vector}
\bibfield{author}{\bibinfo{person}{Pan Xie}, \bibinfo{person}{Qipeng Zhang}, \bibinfo{person}{Zexian Li}, \bibinfo{person}{Hao Tang}, \bibinfo{person}{Yao Du}, {and} \bibinfo{person}{Xiaohui Hu}.} \bibinfo{year}{2022}\natexlab{}.
\newblock \showarticletitle{Vector quantized diffusion model with codeunet for text-to-sign pose sequences generation}.
\newblock \bibinfo{journal}{\emph{arXiv preprint arXiv:2208.09141}} (\bibinfo{year}{2022}).
\newblock


\bibitem[Yang et~al\mbox{.}(2023a)]%
        {yang2023unifiedgesture}
\bibfield{author}{\bibinfo{person}{Sicheng Yang}, \bibinfo{person}{Zilin Wang}, \bibinfo{person}{Zhiyong Wu}, \bibinfo{person}{Minglei Li}, \bibinfo{person}{Zhensong Zhang}, \bibinfo{person}{Qiaochu Huang}, \bibinfo{person}{Lei Hao}, \bibinfo{person}{Songcen Xu}, \bibinfo{person}{Xiaofei Wu}, \bibinfo{person}{Zonghong Dai}, {et~al\mbox{.}}} \bibinfo{year}{2023}\natexlab{a}.
\newblock \showarticletitle{UnifiedGesture: A Unified Gesture Synthesis Model for Multiple Skeletons}.
\newblock \bibinfo{journal}{\emph{arXiv preprint arXiv:2309.07051}}.
\newblock


\bibitem[Yang et~al\mbox{.}(2023b)]%
        {yang2023diffusestylegesture}
\bibfield{author}{\bibinfo{person}{Sicheng Yang}, \bibinfo{person}{Zhiyong Wu}, \bibinfo{person}{Minglei Li}, \bibinfo{person}{Zhensong Zhang}, \bibinfo{person}{Lei Hao}, \bibinfo{person}{Weihong Bao}, \bibinfo{person}{Ming Cheng}, {and} \bibinfo{person}{Long Xiao}.} \bibinfo{year}{2023}\natexlab{b}.
\newblock \showarticletitle{DiffuseStyleGesture: Stylized Audio-Driven Co-Speech Gesture Generation with Diffusion Models}.
\newblock \bibinfo{journal}{\emph{arXiv preprint arXiv:2305.04919}}.
\newblock


\bibitem[Yang et~al\mbox{.}(2024)]%
        {yang2024freetalker}
\bibfield{author}{\bibinfo{person}{Sicheng Yang}, \bibinfo{person}{Zunnan Xu}, \bibinfo{person}{Haiwei Xue}, \bibinfo{person}{Yongkang Cheng}, \bibinfo{person}{Shaoli Huang}, \bibinfo{person}{Mingming Gong}, {and} \bibinfo{person}{Zhiyong Wu}.} \bibinfo{year}{2024}\natexlab{}.
\newblock \showarticletitle{Freetalker: Controllable speech and text-driven gesture generation based on diffusion models for enhanced speaker naturalness}. In \bibinfo{booktitle}{\emph{ICASSP 2024-2024 IEEE International Conference on Acoustics, Speech and Signal Processing (ICASSP)}}. IEEE, \bibinfo{pages}{7945--7949}.
\newblock


\bibitem[Yi et~al\mbox{.}(2023)]%
        {yi2023generating}
\bibfield{author}{\bibinfo{person}{Hongwei Yi}, \bibinfo{person}{Hualin Liang}, \bibinfo{person}{Yifei Liu}, \bibinfo{person}{Qiong Cao}, \bibinfo{person}{Yandong Wen}, \bibinfo{person}{Timo Bolkart}, \bibinfo{person}{Dacheng Tao}, {and} \bibinfo{person}{Michael~J Black}.} \bibinfo{year}{2023}\natexlab{}.
\newblock \showarticletitle{Generating holistic 3d human motion from speech}. In \bibinfo{booktitle}{\emph{Proceedings of the IEEE/CVF Conference on Computer Vision and Pattern Recognition}}. \bibinfo{pages}{469--480}.
\newblock


\bibitem[Yoon et~al\mbox{.}(2020)]%
        {yoon2020speech}
\bibfield{author}{\bibinfo{person}{Youngwoo Yoon}, \bibinfo{person}{Bok Cha}, \bibinfo{person}{Joo-Haeng Lee}, \bibinfo{person}{Minsu Jang}, \bibinfo{person}{Jaeyeon Lee}, \bibinfo{person}{Jaehong Kim}, {and} \bibinfo{person}{Geehyuk Lee}.} \bibinfo{year}{2020}\natexlab{}.
\newblock \showarticletitle{Speech gesture generation from the trimodal context of text, audio, and speaker identity}.
\newblock \bibinfo{journal}{\emph{ACM Transactions on Graphics (TOG)}} \bibinfo{volume}{39}, \bibinfo{number}{6}, \bibinfo{pages}{1--16}.
\newblock


\bibitem[Yoon et~al\mbox{.}(2022)]%
        {yoon2022genea}
\bibfield{author}{\bibinfo{person}{Youngwoo Yoon}, \bibinfo{person}{Pieter Wolfert}, \bibinfo{person}{Taras Kucherenko}, \bibinfo{person}{Carla Viegas}, \bibinfo{person}{Teodor Nikolov}, \bibinfo{person}{Mihail Tsakov}, {and} \bibinfo{person}{Gustav~Eje Henter}.} \bibinfo{year}{2022}\natexlab{}.
\newblock \showarticletitle{The GENEA Challenge 2022: A large evaluation of data-driven co-speech gesture generation}. In \bibinfo{booktitle}{\emph{Proceedings of the 2022 International Conference on Multimodal Interaction}}. \bibinfo{pages}{736--747}.
\newblock


\bibitem[Zhang et~al\mbox{.}(2022)]%
        {zhang2022motiondiffuse}
\bibfield{author}{\bibinfo{person}{Mingyuan Zhang}, \bibinfo{person}{Zhongang Cai}, \bibinfo{person}{Liang Pan}, \bibinfo{person}{Fangzhou Hong}, \bibinfo{person}{Xinying Guo}, \bibinfo{person}{Lei Yang}, {and} \bibinfo{person}{Ziwei Liu}.} \bibinfo{year}{2022}\natexlab{}.
\newblock \showarticletitle{Motiondiffuse: Text-driven human motion generation with diffusion model}.
\newblock \bibinfo{journal}{\emph{arXiv preprint arXiv:2208.15001}}.
\newblock


\bibitem[Zhao et~al\mbox{.}(2023)]%
        {zhao2023modiff}
\bibfield{author}{\bibinfo{person}{Mengyi Zhao}, \bibinfo{person}{Mengyuan Liu}, \bibinfo{person}{Bin Ren}, \bibinfo{person}{Shuling Dai}, {and} \bibinfo{person}{Nicu Sebe}.} \bibinfo{year}{2023}\natexlab{}.
\newblock \showarticletitle{Modiff: Action-Conditioned 3D Motion Generation with Denoising Diffusion Probabilistic Models}.
\newblock \bibinfo{journal}{\emph{arXiv preprint arXiv:2301.03949}} (\bibinfo{year}{2023}).
\newblock


\bibitem[Zhou et~al\mbox{.}(2022)]%
        {zhou2022gesturemaster}
\bibfield{author}{\bibinfo{person}{Chi Zhou}, \bibinfo{person}{Tengyue Bian}, {and} \bibinfo{person}{Kang Chen}.} \bibinfo{year}{2022}\natexlab{}.
\newblock \showarticletitle{Gesturemaster: Graph-based speech-driven gesture generation}. In \bibinfo{booktitle}{\emph{Proceedings of the 2022 International Conference on Multimodal Interaction}}. \bibinfo{pages}{764--770}.
\newblock


\bibitem[Zhu et~al\mbox{.}(2023)]%
        {zhu2023taming}
\bibfield{author}{\bibinfo{person}{Lingting Zhu}, \bibinfo{person}{Xian Liu}, \bibinfo{person}{Xuanyu Liu}, \bibinfo{person}{Rui Qian}, \bibinfo{person}{Ziwei Liu}, {and} \bibinfo{person}{Lequan Yu}.} \bibinfo{year}{2023}\natexlab{}.
\newblock \showarticletitle{Taming Diffusion Models for Audio-Driven Co-Speech Gesture Generation}. In \bibinfo{booktitle}{\emph{Proceedings of the IEEE/CVF Conference on Computer Vision and Pattern Recognition}}. \bibinfo{pages}{10544--10553}.
\newblock


\end{thebibliography}










\end{document}